\documentclass[12pt,a4paper]{article}
\pdfoutput=1
\usepackage{jcappub}
\usepackage{bm}
\usepackage{indentfirst}
\usepackage{amsmath}
\usepackage{graphicx}
\usepackage{float}
\usepackage{amssymb}
\usepackage{subfigure}
\usepackage{hyperref}
\usepackage{array}
\hypersetup{
	colorlinks=true,
	linkcolor=red,
	citecolor=blue,
}

\begin{document}
\title{Detecting electric charge with Extreme Mass Ratio Inspirals}
\author[a]{Chao Zhang,}
\author[a,1]{Yungui Gong\note{Corresponding author.}}
\affiliation[a]{School of Physics, Huazhong University of Science and Technology,
Wuhan, Hubei 430074, China}
\emailAdd{chao\_zhang@hust.edu.cn}
\emailAdd{yggong@hust.edu.cn}

\abstract{
We consider extreme mass ratio inspirals during which an electrically charged compact object with mass $m_p$ and the charge to mass ratio $q$ inspirals around a Schwarzschild black hole of mass $M$.
Using the Teukolsky and generalized Sasaki-Nakamura formalisms for the gravitational and electromagnetic perturbations around a Schwarzschild black hole,
we numerically calculate the energy flux of both gravitational and electromagnetic waves induced by a charged particle moving in circular orbits.
With one year observation of these extreme mass ratio inspirals, we show that space-based gravitational wave detector such as the Laser Interferometer Space Antenna can detect the charge to mass ratio as small as $q\sim 10^{-3}$.}


\maketitle

\section{Introduction}
The first direct detection of gravitational wave (GW) event GW150914 in 2015 by the Laser Interferometer Gravitational-Wave Observatory (LIGO) Scientific Collaboration and the Virgo Collaboration \cite{Abbott:2016blz,TheLIGOScientific:2016agk} provides us a new window to understand the nature of gravity
in the nonlinear and strong field regimes.
To date, there have been tens of confirmed GW detections in the frequency range of tens to hundreds Hertz \cite{LIGOScientific:2018mvr,LIGOScientific:2020ibl,LIGOScientific:2021usb,LIGOScientific:2021djp}.
However, due to the seismic noise and gravity gradient noise, the ground-based GW observatories can only measure GWs within the frequency range $10-10^3$ Hz, and are unable to explore lower frequency band $10^{-4}-10^{-1}$ Hz where a wealth of astrophysical signals reside.
The proposed space-based observatories like the Laser Interferometer Space Antenna (LISA) \cite{Danzmann:1997hm,Audley:2017drz},
TianQin \cite{Luo:2015ght} and Taiji \cite{Hu:2017mde}
are expected to detect GWs in this low-frequency regime.
Extreme mass ratio inspirals (EMRIs) are among the most promising GW sources expected to be detected for these future space-based GW detectors.
An EMRI consists of a stellar mass compact object (secondary object) with mass $m_p$ such as a black hole (BH) or a neutron star orbiting around a supermassive black hole (SMBH) (primary object) of mass $M$, with the mass ratio $m_p/M$ in the range of $10^{-7}-10^{-4}$.
The binary emits GWs while the secondary object slowly inspirals into the primary object under the gravitational radiation reaction, so EMRIs provide a unique wealth of information about the masses, spins, electric charges, the strong-field physics in the vicinity of BHs, and the astrophysics of their stellar environments, etc \cite{Amaro-Seoane:2007osp,Babak:2017tow,Berry:2019wgg,Fan:2020zhy,Zi:2021pdp}.

The existence of stable charged astrophysical compact objects in nature remains controversial.
Glendening \cite{Glendenning} pointed out that macroscopic bodies can have a small amount of charge of approximately $100$ Coulombs per solar mass, which does not affect much on the structure of the star.
In such a system that the star is modeled by a ball of hot ionized gas, the electrons are likely to rise to the top and escape from the star because of the difference in the partial pressure of electrons compared to that of ions.
The electric field will be created by the charge separation and then prevent the electrons from escaping further from the star.
The electrified star can have positive charge of about 100 Coulombs per solar mass when the equilibrium is reached.
However, for highly relativistic stars, the situation might be different when we consider the full effects of general relativity.
The highly compact stars, whose radius is on the verge of forming an event horizon, can contain a maximal huge amount of charge with the charge to mass ratio of the order one \cite{Bekenstein:1971ej,de1995relativistic,de1999relativistic,PhysRevD.65.104001,PhysRev.72.390,zhang_influence_1982,Anninos:2001yb,bonnor1975very}.
The huge amount of charge supports the global balance of forces between the matter part and the electrostatic part.
Nevertheless, this balance is unstable and the charged star will collapse to a charged black hole with a perturbation of decrease in the electric field \cite{Ray:2003gt}.
The electric charge can also play a vital role in halting gravitational collapse through the equilibrium of large bodies \cite{bonnor1965equilibrium}.
The charged parameter can be applied to construct models of various astrophysical objects of immense gravity by considering the relevant matter distributions.
Such models with charged parameter can successfully describe the characteristics of compact stellar objects like neutron stars, quark stars, etc \cite{Takisa:2014sva}.
Direct observation of electrically charged compact bodies would be very challenging, because of the weak electromagnetic signal and absorption of interstellar dust.
Nonetheless, EMRIs can be excellent system for detecting electrically charged compact bodies.
The extra emission channel due to electromagnetic radiation accelerates the coalescence,
leading to a cumulative dephasing of GWs, which can be distinguished by future space-based GW detectors like LISA, TianQin and Taiji.

The gravitational and electromagnetic radiations from binary BHs with electric and magnetic charges were studied in \cite{Liu:2020vsy}.
The authors studied the motion of binary BHs with both electric and magnetic charges on circular orbits by combining the Newtonian gravity with the radiation reaction.
They found that electric and magnetic charges significantly suppress the merger time of dyonic binaries.
The work was later generalized to eccentric orbits and they derived the waveforms for dyonic binary BH inspirals \cite{Liu:2020bag}.
In \cite{Cardoso:2020iji}, the authors took into account of astrophysical environments (e.g., accretion disks) in addition to the scalar/vector and gravitational radiations to
investigate the eccentricity and orbital evolution for BH binaries, and discussed the  competition between radiative mechanisms and environmental effects on the eccentricity evolution.
The effect of charged BHs with electromagnetic dipole emission on the estimation of the chirp mass and BH merger were studied in \cite{Christiansen:2020pnv,Jai-akson:2017ldo}.
The merger rate distribution of primordial black hole binaries with electric charges and magnetic charges was studies in \cite{Liu:2022wtq,Liu:2020cds}.
However, these discussions are mainly based on the Newtonian orbit and dipole emission.
When the secondary object approaches the innermost stable circular orbit (ISCO) of the central BH, the Newtonian approximation breaks down and the above method  becomes ill suited in the late phase of inspiral.
To use the highly relativistic dynamics of EMRIs with charged secondary object to place more stringent constraint on the possible existed electric charge,
we need to get more accurate orbital motion and gravitational waveforms.
In this paper, we use the Teukolsky formalism for BH perturbations \cite{Teukolsky:1973ha,Press:1973zz,Teukolsky:1974yv,Zerilli:1970se,Chandrasekhar:1975nkd}
to calculate the orbital evolution of inspiralling into the ISCO.
A more detailed presentation of the Teukolsky formalism can be found in \cite{Poisson:1993vp,Poisson:1995vs,Apostolatos:1993nu,Cutler:1994pb}.

The paper is organized as follows.
In Sec. \ref{sec2}, we introduce the Einstein-Maxwell field equations, the Teukolsky perturbation formalism, and the source terms for the inhomogeneous Teukolsky equations.
In Sec. \ref{sec3}, we numerically calculate the energy flux for gravitational and electromagnetic field using the Teukolsky and generalized Sasaki-Nakamura formalisms.
In Sec. \ref{sec4}, we give the numerical results of energy flux and use the dephasing of GW to constrain the charge to mass ratio.
Sec. \ref{sec5} is devoted to conclusions and discussions.

\section{Einstein-Maxwell field equations}
\label{sec2}
The Einstein equations and Maxwell equations are
\begin{equation}
G^{\mu\nu}=8\pi T^{\mu\nu}_p+8\pi T^{\mu\nu}_e,
\end{equation}
\begin{equation}
\nabla_\nu F^{\mu\nu}=4\pi J^\mu,
\end{equation}
where $G^{\mu\nu}$ is Einstein tensor, $F_{\mu\nu}=\nabla_\mu A_\nu-\nabla_\nu A_\mu$ is the electromagnetic field tensor, $J^\mu$ is the electric 4-current density, $T_p^{\mu\nu}$ and $T_e^{\mu\nu}$ are the particle's stress-energy tensor and electromagnetic stress-energy tensor, respectively.
We consider a charged body with mass $m_p$ and electric charge $q$ per mass (charge to mass ratio) orbiting around a BH of mass $M$.
For an EMRI system with $m_p\ll M$, we can ignore the contribution to the background metric from the electromagnetic field.
Since the amplitude of the electromagnetic stress-energy $T_e^{\mu\nu}$ is quadratic in the electromagnetic field, the contribution to the background metric from the electromagnetic field is second order.
The perturbed Einstein and Maxwell equations for EMRIs are
\begin{equation}
G^{\mu\nu}=8\pi T^{\mu\nu}_p,
\end{equation}
\begin{equation}
\nabla_\nu F^{\mu\nu}=4\pi J^\mu,
\end{equation}
where
\begin{equation}
T^{\mu\nu}_p(x)=m_p\int d\tau~ u^{\mu}u^\nu\delta^{(4)}\left[x-z(\tau)\right],
\end{equation}
\begin{equation}
J^{\mu}(x)=q m_p\int d\tau~ u^{\mu}\delta^{(4)}\left[x-z(\tau)\right].
\end{equation}

We apply the Newman-Penrose formalism \cite{Newman:1966ub} to discuss the perturbations induced by a charged particle with mass $m_p$ and electric charge $q$ per mass around a Schwarzschild BH.
The metric of Schwarzchild BHs is
\begin{equation}\label{SBH}
ds^2= f(r)dt^2-\frac{1}{f(r)}dr^2-r^2d\Omega^2,
\end{equation}
where $f(r)=1-2M/r$.
Based on the metric \eqref{SBH}, we construct the null tetrad,
\begin{equation}
\begin{split}
l^\mu&=\left[ \frac{r}{r-2M}, 1,0,0 \right],\\
n^\mu&=\left[\frac{1}{2} , -\frac{r-2M}{2r}, 0, 0 \right],\\
m^\mu&=\left[0 , 0,\frac{1}{\sqrt{2}r} ,\frac{i}{\sqrt{2}r\sin\theta} \right],\\
\bar{m}^\mu&=\left[0 , 0,\frac{1}{\sqrt{2}r} ,\frac{-i}{\sqrt{2}r\sin\theta} \right].
\end{split}
\end{equation}
Then the propagating electromagnetic field is described by the two complex quantities,
\begin{equation}
\phi_0=F_{\mu\nu}l^\mu m^\nu,\qquad \phi_2=F_{\mu\nu}\bar{m}^\mu n^\nu.
\end{equation}
The propagating gravitational field is described by the two complex Newman-Penrose variables
\begin{equation}
\psi_0=-C_{\alpha\beta\gamma\delta}l^\alpha m^\beta l^\gamma m^\delta    ,\qquad \psi_4=-C_{\alpha\beta\gamma\delta}n^\alpha \bar{m}^\beta n^\gamma \bar{m}^\delta,
\end{equation}
where $C_{\alpha\beta\gamma\delta}$ is the Weyl tensor.
A single master equation for a gravitational perturbation ($s=-2$) and an electromagnetic field ($s=-1$) was derived in Ref.~\cite{Teukolsky:1973ha},
\begin{equation}
\label{TB}
\begin{split}
 & \frac{r^2}{f(r)}\frac{\partial^2 \psi}{\partial t^2}-\frac{1}{\sin^2\theta}\frac{\partial^2 \psi}{\partial \varphi^2}-\Delta^{-s}\frac{\partial}{\partial r}\left(\Delta^{s+1}\frac{\partial \psi}{\partial r}\right)   -\frac{1}{\sin \theta}\frac{\partial}{\partial \theta}\left(\sin\theta\frac{\partial \psi}{\partial \theta}\right)    \\
 &\qquad-\frac{2i s\cos\theta}{\sin^2\theta}\frac{\partial \psi}{\partial \varphi}-2s \left[\frac{M r^2}{\Delta}-r\right]\frac{\partial \psi}{\partial t}+(s^2\cot^2\theta-s)\psi=4\pi r^2 T,
\end{split}
\end{equation}
where $\Delta=r^2-2M r$,
the explicit field $\psi$ and the corresponding source $T$ are given in  Table \ref{source} \cite{Teukolsky:1973ha}.
 \begin{table}[h]
  \centering
  	\begin{tabular}{|p{1cm}<{\centering}|p{2cm}<{\centering}|p{2cm}<{\centering}|}
		\hline
$s$ & $\psi$ &  $T$\\ \hline
-1   & $r^2\phi_2$ & $r^2J_2$  \\ \hline
 -2   & $r^4\psi_4$ &$2r^4T_4$  \\ \hline
	\end{tabular}
 \caption{The explicit expressions for the the field $\psi$ and the corresponding source $T$.}
    \label{source}
\end{table}
In terms of the spin-weighted spherical harmonics ${_{s}}Y_{lm}(\theta,\varphi)$ \cite{Goldberg:1966uu}, the field $\psi$ can be written as
\begin{equation}
\psi=\int d\omega \sum_{l,m}R_{\omega lm}(r)~{_{s}}Y_{lm}(\theta,\varphi)e^{-i\omega t},
\end{equation}
where the radial function $R_{\omega lm}(r)$ satisfies the inhomogeneous Teukolsky equation
\begin{equation}
\label{Teukolsky}
\Delta^{-s}\frac{d}{d r}\left(\Delta^{s+1}\frac{d R_{\omega lm}}{d r}\right)+\left(\frac{\omega^2r^4-2i s\omega r^2 (r-M)}{\Delta} +4is \omega r-\lambda    \right)R_{\omega lm}=T_{\omega lm},
\end{equation}
$\lambda=(l-s)(l+s+1)$, and the source $T_{\omega lm}(r)$ is
\begin{equation}
T_{\omega lm}(r)=\frac{1}{2\pi}\int dt d\Omega ~4\pi r^2 T ~{_s}\bar{Y}_{lm}(\theta,\varphi)e^{i\omega t}.
\end{equation}
For a circular trajectory at $r_0$ under consideration
\begin{equation}
\begin{split}
T^{\mu\nu}_p(x)&=\frac{m_p}{r_0^2}\frac{u^{\mu}u^{\nu}}{u^t}\delta(r-r_0)\delta(\cos\theta)\delta(\varphi-\Omega t),\\
J^{\mu}(x)&=q\frac{m_p}{r_0^2}\frac{u^{\mu}}{u^t}\delta(r-r_0)\delta(\cos\theta)\delta(\varphi-\Omega t),
\end{split}
\end{equation}
where
\begin{equation}
u^{\mu}=\left(\tilde{E}/f_0,0,0,\tilde{L}/r_0^2\right),
\end{equation}
the particle's energy per mass $\tilde{E}$, angular momentum per mass $\tilde{L}$, and angular velocity $\Omega$ are \cite{Poisson:1993vp}
\begin{equation}
\begin{split}
\tilde{E}&=f_0(1-3M/r_0)^{-1/2},  \\
\tilde{L}&=(Mr_0)^{1/2}(1-3M/r_0)^{-1/2}, \\
\Omega&=(M/r_0^3)^{-1/2},
\end{split}
\end{equation}
with $f_0=1-2M/r_0$.
The source term $T_{\omega lm}(r)$ for gravitational perturbation $(s=-2)$ was  given in \cite{SASAKI198185}, 
while the source term for the electromagnetic field perturbation $(s=-1)$ reads
\begin{equation}
\begin{split}
T_{\omega lm}(r)=&-i\sqrt{2\frac{r}{M}} \pi q \frac{m_p}{M}(\frac{r}{M}-2) ~{_{-1}}Y_{lm}\left(\frac{\pi}{2},0\right)\delta'(r-r_0) \delta(\omega-m\Omega)\\
&-\pi  q \frac{m_p}{M} \left(2- \frac{r}{M}  \right)\sqrt{2l (l+1)}~{_{0}}Y_{lm}\left(\frac{\pi}{2},0\right)\delta(r-r_0)\delta(\omega-m\Omega)\\
&-\frac{\pi  q m_p \left(-2 r^2 \omega  +i r M-2 i M^2 \right)}{M^{5/2}\sqrt{2r}}~{_{-1}}Y_{lm}\left(\frac{\pi}{2},0\right)\delta(r-r_0)\delta(\omega-m\Omega).
\end{split}
\end{equation}

\section{The numerical calculation for energy flux}
\label{sec3}
The homogeneous Teukolsky equation \eqref{Teukolsky} admits two linearly independent solutions $R^{\text{in}}_{\omega lm}$ and $R^{\text{up}}_{\omega lm}$, with the following asymptotic values at the horizon $r=2M$ and at infinity,
\begin{equation}
R^{\text{in}}_{\omega lm}=
\begin{cases}
\Delta^{-s}e^{-i\omega r^*},&\qquad (r^*\to-\infty)\\
B^{\text{out}}\frac{e^{i\omega r^*}}{r^{2s+1}}+B^{\text{in}}\frac{e^{-i\omega r^*}}{r}, &\qquad (r^*\to+\infty)
\end{cases}
\end{equation}
\begin{equation}
R^{\text{up}}_{\omega lm}=
\begin{cases}
D^{\text{up}}e^{i\omega r^*}+D^{\text{in}}\Delta^{-s}e^{-i\omega r^*},&\qquad (r^*\to-\infty)\\
e^{i\omega r^*},&\qquad (r^*\to+\infty)\\
\end{cases}
\end{equation}
where the tortoise radius $r^*=r+2M\ln{\left(r/2M-1\right)}$.
The solutions $R^{\text{in}}_{\omega lm}$ and $R^{\text{up}}_{\omega lm}$  are purely outgoing at infinity and purely ingoing at the horizon.
With the help of these homogeneous solutions, the solution to Eq.~\eqref{Teukolsky} is
\begin{equation}
R_{\omega lm}(r)=\frac{R^{\text{in}}_{\omega lm}\int_{r^*}^{+\infty}\Delta^{s}R^{\text{up}}_{\omega lm}T_{\omega lm}dr^*+R^{\text{up}}_{\omega lm}\int_{-\infty}^{r^*}\Delta^{s}R^{\text{in}}_{\omega lm}T_{\omega lm}dr^*}{2i\omega B^{\text{in}}}.
\end{equation}
The energy per unit solid angle carried off by outgoing electromagnetic waves at infinity is
\begin{equation}
\frac{d^2 E^{\infty}_q}{dtd\Omega}=\lim_{r\to\infty}\frac{r^2}{2\pi}|\phi_2|^2.
\end{equation}
Since
\begin{equation}
R_{\omega lm}(r\to\infty)=r^2\phi_2=q Z^{\infty}_{\omega lm}\delta(\omega-m\Omega)re^{i\omega r^*},
\end{equation}
so the expression for the energy flux of electromagnetic waves is
\begin{equation}
\label{einfty}
dE_q^{\infty}/dt=\dot{E}_q^{\infty}=q^2\left(\frac{m_p}{M}\right)^2\sum_{l=1}^{\infty}\sum_{m=1}^{l} \frac{|Z^\infty_{\omega lm}|^2}{\pi},
\end{equation}
where
\begin{equation}
\begin{split}
Z^\infty_{\omega lm}=&\frac{1}{2i\omega qB^{\text{in}}}\int_{2M}^{\infty}dr'\frac{R^\text{in} T_{\omega lm}(r)}{\Delta} \\
=&\frac{1}{2i\omega  B^{\text{in}}}\left\{  \frac{i \sqrt{2} \pi   ~{_{-1}}Y_{lm}\left(\frac{\pi}{2},0\right) }{\sqrt{r/M}}\frac{d R^\text{in}_{\omega lm}}{dr}\right.\\
&\left.+\sqrt{2} \pi\left(\frac{   \sqrt{l (l+1)}}{r/M}~{_{0}}Y_{lm}\left(\frac{\pi}{2},0\right) +\frac{   \left(r^2 \omega -iM r+2 iM^2\right)}{M^{-1/2}(r-2M) r^{3/2}} {_{-1}}Y_{lm}\left(\frac{\pi}{2},0\right)\right) R^\text{in}_{\omega lm} \right\}.
\end{split}
\end{equation}
The energy carried off by ingoing electromagnetic waves at the horizon is
\begin{equation}
\label{ehorison}
dE_q^{H}/dt=\dot{E}_q^{H}=q^2\left(\frac{m_p}{M}\right)^2\sum_{l=1}^{\infty}\sum_{m=1}^{l} \frac{|Z^H_{\omega lm}|^2}{\pi},
\end{equation}
where
\begin{equation}
\begin{split}
Z^H_{\omega lm}=&\frac{1}{2i\omega  B^{\text{in}}}\left\{  \frac{i \sqrt{2} \pi   ~{_{-1}}Y_{lm}\left(\frac{\pi}{2},0\right) }{\sqrt{r/M}}\frac{d R^\text{up}_{\omega lm}}{dr}\right.\\
&\left.+\sqrt{2} \pi\left(\frac{   \sqrt{l (l+1)}}{r/M}~{_{0}}Y_{lm}\left(\frac{\pi}{2},0\right) +\frac{  \left(r^2 \omega -iM r+2 i M^2\right)}{M^{-1/2}(r-2M) r^{3/2}} {_{-1}}Y_{lm}\left(\frac{\pi}{2},0\right)\right) R^\text{up}_{\omega lm}\right\}.
\end{split}
\end{equation}
The total energy flux emitted by electromagnetic fields is
\begin{equation}
\dot{E}_q=    \dot{E}^\infty_q+\dot{E}^H_q.
\end{equation}
The total energy flux emitted by tensor fields is
\begin{equation}
\dot{E}_{t}=\dot{E}^\infty_{t}+\dot{E}^H_{t},
\end{equation}
where the formula for $\dot{E}_{t}^{\infty,H}$ is given in Refs. \cite{Poisson:1993vp,Cutler:1994pb}.

The generalized Regge-Wheeler equation is
\begin{equation}
\label{RW}
\left[\frac{d^2}{dr^{*2}}+\omega^2-V_{\text{RW}}(r,s)\right]X_{\omega lm}  =0,
\end{equation}
where
\begin{equation}
V_{\text{RW}}(r,s)=f(r)\left[\frac{l(l+1)}{r^2}-\frac{2(s^2-1)}{r^3}\right].
\end{equation}
The homogeneous Regge-Wheeler equation \eqref{RW} admits two linearly independent solutions $X^{\text{in}}_{\omega lm}$ and $X^{\text{up}}_{\omega lm}$ which are implemented in the Mathematica packages of the BH Perturbation Toolkit \cite{BHPToolkit}, with the following asymptotic values at the horizon $r=2M$ and at infinity,
\begin{equation}
X^{\text{in}}_{\omega lm}=
\begin{cases}
e^{-i\omega r^*},&\qquad (r^*\to-\infty)\\
A^{\text{out}}e^{i\omega r^*}+A^{\text{in}}e^{-i\omega r^*}, &\qquad (r^*\to+\infty)
\end{cases}
\end{equation}
\begin{equation}
X^{\text{up}}_{\omega lm}=
\begin{cases}
C^{\text{up}}e^{i\omega r^*}+C^{\text{in}}e^{-i\omega r^*},&\qquad (r^*\to-\infty)\\
e^{i\omega r^*}.&\qquad (r^*\to+\infty)\\
\end{cases}
\end{equation}
The solutions $X^{\text{in}}_{\omega lm}$ and $X^{\text{up}}_{\omega lm}$  are purely outgoing at infinity and purely ingoing at the horizon.
The Wronskian of $X^{\text{in}}_{\omega lm}$ and $X^{\text{up}}_{\omega lm}$ is
\begin{equation}
W=X^{\text{in}}_{\omega lm}\frac{dX^{\text{up}}_{\omega lm}}{dr^*}-X^{\text{up}}_{\omega lm}\frac{dX^{\text{in}}_{\omega lm}}{dr^*}=2i\omega A^{\text{in}}.
\end{equation}
The relation between $B^{\text{in}}$ and $A^{\text{in}}$ for $s=-2$ is given in \cite{Poisson:1993vp}, while for $s=-1$
\begin{equation}
B^{\text{in}}=\frac{-l(l+1)A^{\text{in}}}{2i\omega}.
\end{equation}
With the solutions $X^{\text{in},\text{up}}_{\omega lm}$,
we can obtain $R^{\text{in},\text{up}}_{\omega lm}$ as follows \cite{Hughes:2000pf},
\begin{equation}
R^{\text{in},\text{up}}_{\omega lm}=\left(\frac{\Delta}{r}\right)^{|s|}\mathcal{L}^{|s|}\left[\left(\frac{r}{\sqrt{\Delta}}\right)^{|s|}\frac{X^{\text{in},\text{up}}_{\omega lm}}{r\sqrt{\Delta^s}}\right],
\end{equation}
where $\mathcal{L}=d/dr+i\omega/f(r)$.
\section{Results}
\label{sec4}
\begin{figure}
    \centering
    \includegraphics[width=0.9\columnwidth]{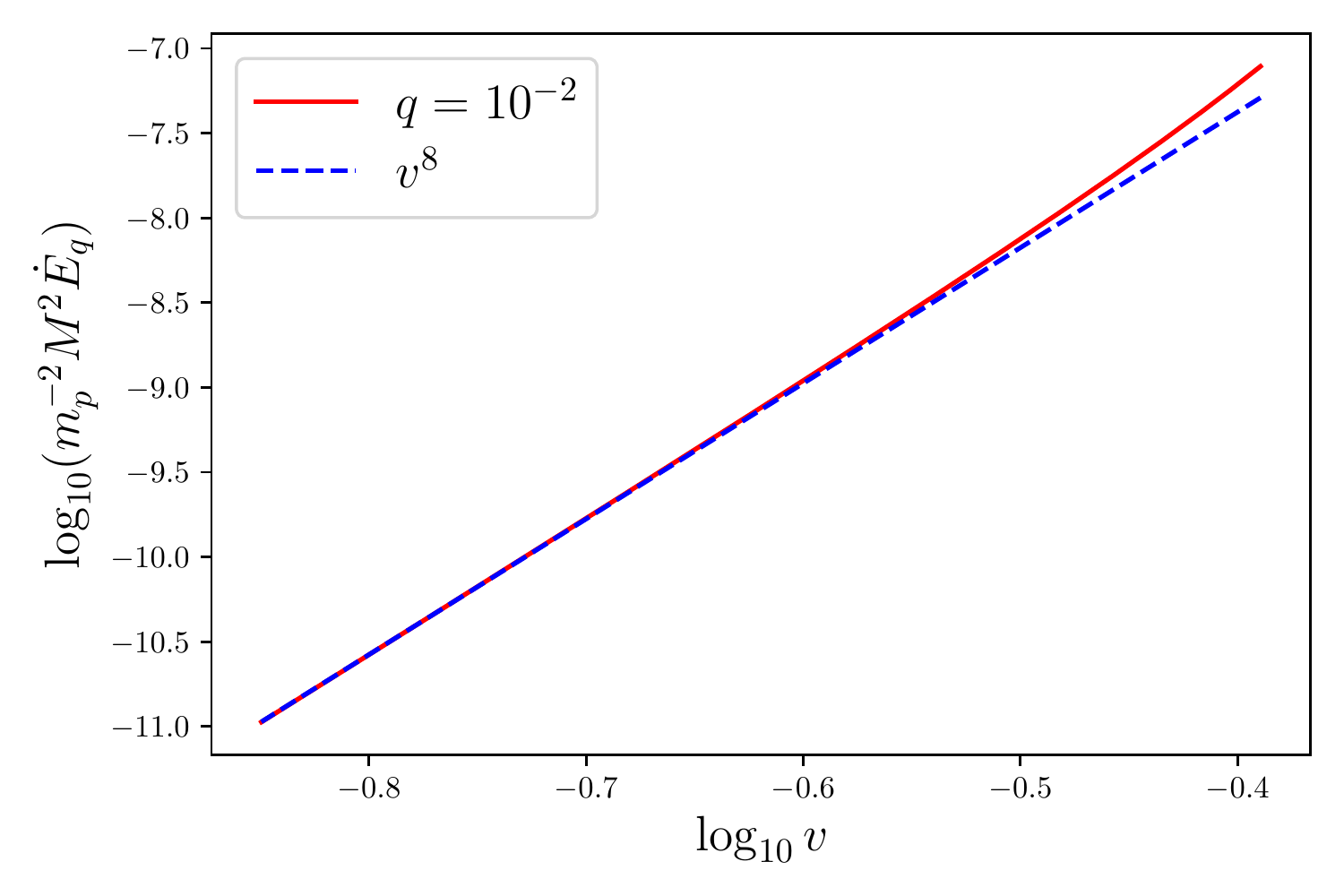}
    \caption{The radiation power from charged particle with the charge to mass ratio $q=10^{-2}$ as a function of the orbital speed $v=(M\Omega)^{1/3}$.
    The blue dashed line represents the emission power $v^8$ due to electromagnetic dipole radiation in a Newtonian binary or flat spacetime background.}
    \label{energy}
\end{figure}

\begin{figure}
    \centering
    \includegraphics[width=0.9\columnwidth]{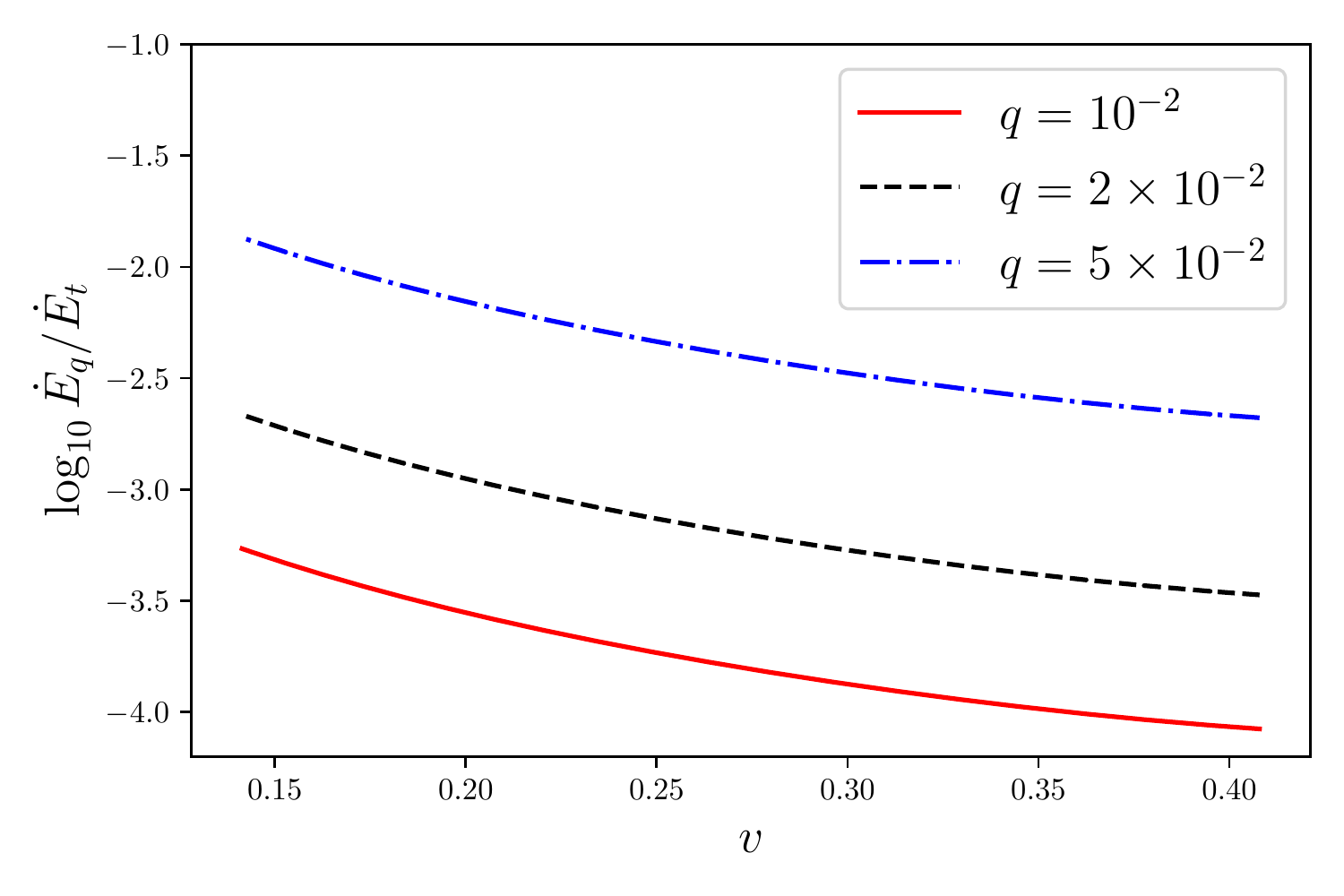}
    \caption{The relative difference between the GW flux and the electromagnetic flux as a function of the orbital speed $v=(M\Omega)^{1/3}$ for a charged particle with different charge to mass ratio $q$.}
    \label{eratio}
\end{figure}

From Eqs.~\eqref{einfty} and \eqref{ehorison}, the energy flux $\dot{E}_q$ emitted by electromagnetic field is proportional to the squared mass ratio $(m_p/M)^2$ and squared electric charge $q^2$.
Fig. \ref{energy} shows the normalized flux $m_p^{-2}M^2\dot{E}_q$ which  depends only on the dimensionless electric charge $q$ (the charge to mass ratio).
The emission power due to electromagnetic radiation increases as the orbital velocity $v=(M\Omega)^{1/3}$ increases while the particle inspiraling toward the ISCO at $r=6~M$.
The extra emission channel due to electromagnetic radiation accelerates the coalescence, leaving a significant imprint on gravitational waves.
From Fig.~\ref{energy}, we see that the behavior of electromagnetic flux in the case of slow motion with the velocity $v\ll 1$ is consistent with the result of electromagnetic dipole radiation for a Newtonian binary \cite{Liu:2020cds,Liu:2020vsy,Cardoso:2020iji}
\begin{equation}
 \dot{E}_q^{\text{Newtonian}}=\frac{2q^2m_p^2M^2}{3r^4}=\frac{2q^2m_p^2}{3M^2}v^8.
\end{equation}
Far away from the central BH, the speed of the particle is much less than the speed of light,
and the main electromagnetic radiation is the electric dipole radiation with the radiated power falling off as $v^8$.
In the strong field regions near ISCO, the emission power of electromagnetic waves differ from $v^8$,
this can be interpreted as the interaction between electromagnetic filed and gravitational field.
To compare electromagnetic and gravitational radiation,
we show the ratio $\dot{E}_q/\dot{E}_t$ as a function of the orbital speed $v$ in Fig.~\ref{eratio}.
The ratio decreases as the orbital speed increases, since the emission power $\dot{E}_t$ of gravitational waves grows as $v^{10}$ \cite{Apostolatos:1993nu}, so it grows faster than the that of electromagnetic waves with $v^{8}$ on the assumption that weak field limit is still applicable.
As expected in Fig. \ref{eratio}, the contribution of electromagnetic waves becomes larger if the particle has more charge.

\begin{figure}
    \centering
    \includegraphics[width=0.9\columnwidth]{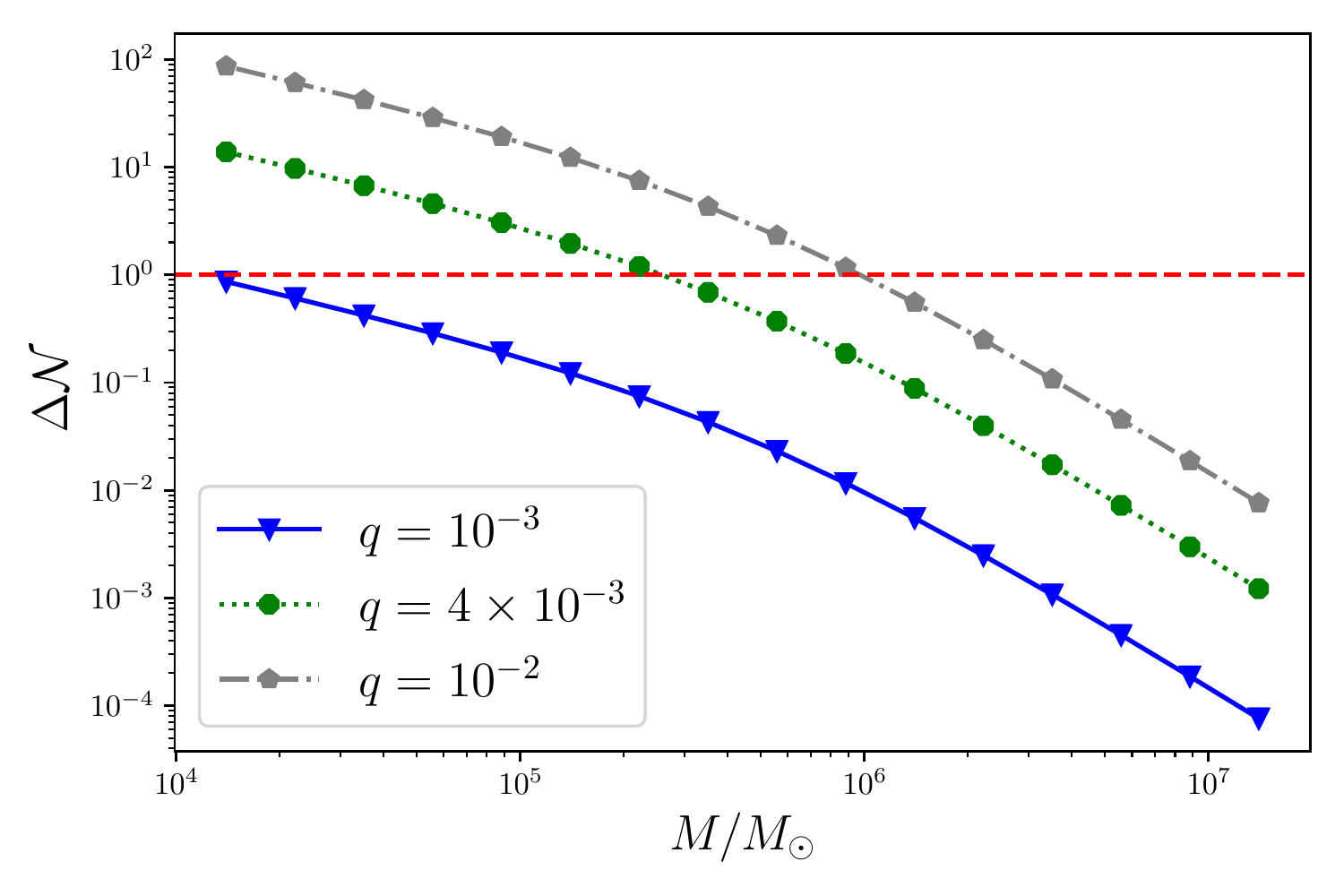}
    \caption{The difference in the number of GW cycles accumulated by EMRIs on circular orbit. The mass of the charged particle is fixed to be $m_p=1.4~M_{\odot}$, the mass of the central black hole varies in the range $M\in\left[1.4\times 10^4,1.4\times 10^7\right]~M_\odot$, and the charge to mass ratio $q$ is chosen as $q=10^{-3}$ (blue solid line), $q=4\times 10^{-3}$ (green dotted line), $q=10^{-2}$ (grey dash-dotted line), respectively. The red dashed line corresponds to the threshold above which the two signals are distinguished by LISA. All binaries are observed one year before merger.}
    \label{phase}
\end{figure}

To detect the electric charge of the small compact object (particle) in the EMRI, 
we compare the  number of cycles between charged and uncharged particle accumulated before the merger \cite{Berti:2004bd},
\begin{equation}
\mathcal{N}=\int_{f_{\text{min}}}^{f_{\text{max}}} \frac{f_{\text{GW}}}{\dot{f}_{\text{GW}}}df_{\text{GW}},
\end{equation}
where the GW frequency $f_{\text{GW}}=2\Omega$,
\begin{equation}
\label{evolution}
\dot{f}_{\text{GW}}=\frac{3}{2}\frac{f_{\text{GW}}}{r_0}\left.\frac{dr}{d \tilde{E}}\right|_{r_0} \dot{E},
\end{equation}
and $\tilde{E}$ is the particle's orbital energy per mass.
We choose  $f_{\text{max}}=(6^{3/2}\pi M)^{-1}$ and $f_{\text{min}}=\text{max}\left[f_T,10^{-4}\right]$,
where $f_T$ is the GW frequency one year before the ISCO \cite{Pani:2011xj}
and the cutoff frequency for LISA is $10^{-4}$ \cite{LISA:2017pwj}.
In the Newtonian approximation, $\tilde{E}=-M/(2r)$, Eq.~\eqref{evolution} reduces to the major semiaxis evolution in \cite{Cardoso:2020iji},
\begin{equation}
\dot{r}=-\frac{2r^2}{Mm_p}\dot{E}.
\end{equation}
Now we fix the particle's mass to be $m_p=1.4~M_{\odot}$, but vary its charge $q$ and choose different mass $M\in\left[1.4\times 10^4,1.4\times 10^7\right]~M_\odot$ for the central black hole to calculate $\Delta\mathcal{N}=\mathcal{N}(q=0)-\mathcal{N}(q)$ and the result is shown in
Fig. \ref{phase}.
As discussed above,
the extra emission channel due to electromagnetic radiation accelerates the coalescence,
so for the same one year observation before the merger,
the charged particle starts further away from ISCO and the difference $\Delta\mathcal{N}$ is always positive.
As shown in Fig. \ref{phase}, $\Delta \mathcal{N}$ increases monotonically as the charge to mass ratio $q$ becomes bigger, and it strongly depends on the mass of the central BH such that lighter BHs have larger $\Delta \mathcal{N}$.
Following Ref. \cite{Maselli:2020zgv}, we take the threshold for a detectable dephasing that two signals are distinguished by LISA as 1 radian.
From Fig.~\ref{phase}, we see that the dephasing $\Delta\mathcal{N}$ can be larger than 1 radian  for $M\sim 10^4~M_{\odot}$ and $q\geq 10^{-3}$ with one year's observation before merger.
Therefore LISA can detect the charge to mass ratio $q\sim10^{-3}$ with one year observation of EMRIs.

\section{Conclusions}
\label{sec5}
We study the energy emissions and GWs from EMRIs with a small charged compact object with mass $m_p$ and the charge to mass ratio $q$ inspiraling into a SMBH.
We derive the source term $T_{\omega lm}$ for solving the inhomogeneous Teukolsky equation with electromagnetic field, and calculate the total emission power due to both gravitational and electromagnetic radiations.
Compared with the power emitted by electromagnetic field from the circular orbit in flat background as shown in Fig. \ref{energy}, the emission power in strong gravitational field is bigger for the reason of the coupling between gravitational and electromagnetic fields.
We also discuss the ratio $\dot{E}_q/\dot{E}_t$ as a function of the orbital speed $v$,
and we find that the gravitational radiation $\dot{E}_t$ grows faster than the electromagnetic field contribution $\dot{E}_q$  as the orbital speed $v$ increases.
By using the difference of number of cycles $\Delta\mathcal{N}$ accumulated before the merger between a charged particle with the charge to mass ratio $q$ and a neutral particle,
we demonstrate that LISA can detect the charge to mass ratio of the small compact object in an EMRI on circular orbits as small as $10^{-3}$.
Our method of calculation can be straightforwardly  extended to generic orbits around Kerr BHs.

\begin{acknowledgments}
We thank Ning Dai and Hong Guo for useful discussions.
This work makes use of the Black Hole Perturbation Toolkit package.
The numerical computations were performed at the public computing service platform provided by Network and Computing Center of HUST.
This research is supported in part by the National Natural Science
Foundation of China under Grant No. 11875136.
\end{acknowledgments}


\providecommand{\href}[2]{#2}\begingroup\raggedright\endgroup

\end{document}